\documentclass{PoS}
\usepackage{epsfig}

\title{The static quark potential for dynamical domain wall fermion simulations}

\ShortTitle{The static quark potential for DWF simulations}

\author{\speaker{Min Li}, for the RBC and UKQCD Collaborations\\
        Department of Physics,Columbia University, New York, NY, 10027, USA\\
        E-mail: \email{minxolee@phys.columbia.edu}\\
        }

\abstract{We present preliminary results for the static quark potential
computed on some of the DWF lattice configurations generated by
the RBC-UKQCD collaborations. Most of these results were obtained
using Wilson lines joining spatial planes fixed into the Coulomb
gauge. We compare the results from this method with the earlier 
ones on $16^3 \times 32$ lattices using Bresenham spatial paths 
with APE smeared link variables. Some preliminary results on $24^3 \times 64$ 
lattices are also presented.
}

\FullConference{XXIVth International Symposium on Lattice Field Theory\\
                July 23-28, 2006\\
                Tucson, Arizona, USA}

\begin{document}

\section{Introduction}
The lattice scale is a very important quantity for lattice QCD,
especially as we work toward more accurate results. Different from
the methods of using hadronic masses and matrix elements to
determine the lattice scale, the static quark potential provides
an independent method. In this proceeding I will provide a status
report on the RBC and UKQCD collaborations' efforts to calculate
the potential on $N_f=2+1$ domain wall fermion dynamical lattice
configurations, and the preliminary results of the corresponding lattice 
spacing assuming $r_0=0.5fm$. The proceeding is organized 
as the following: first I start from the basic mathematical formulation
of potential calculation, and introduce 2 different methods, Coulomb 
gauge method and Bresenham method. After explaining how to use the 
fitting methods to get the lattice spacing I will focus on the 3 specific 
ensembles of lattice configurations, and talk about the results and the
underlying difficulties. Finally I conclude that the Coulomb method is
much easier to implement and is reliable on large volume lattices.

\section{The static quark potential and lattice scale}
The static quark potential between infinitely heavy quark and
anti-quark separated by $\vec{r}$ is calculated from the Wilson
loop: $\left\langle
W(\vec{r},t)\right\rangle=C(\vec{r})e^{-V(\vec{r})t}+(\mbox{"excited
states"})$. Among the various ways to calculate the Wilson loop,
 we focus on 2 here: Bresenham method and Coulomb gauge method. 
While some smearing should be done on the spacial links 
in order to improve the signal/noise ratio, each method will 
implement the smearing in its own way.

The Bresenham method involves APE smearing for
spatial links. While the on axis Wilson loops can be trivially 
calculated using the definition, for the off axis loops, one can 
implement the Bresenham algorithm~\cite{Bresenham} to approximate 
the diagonal path by the closest link path. 
This method was studied in detail in Ref.~\cite{koichi}.

The Coulomb gauge method was first introduced in Ref.~\cite{Coulomb},
The main idea is to gauge fix the lattices into the Coulomb gauge
on each 3-dimensional plane orthogonal to the time direction, and 
then just compute the trace of the product of pairs of temporal links
whose ends are fixed in Coulomb gauge. 

In many respects, the two methods are quite different and have their
own pros and cons. The Bresenham method uses APE smearing. By adjusting
the smearing to maximize the overlap to the ground state it can safely 
eliminate most of the excited states contamination. On the other hand, 
the complexity involved in the various spacial paths makes the method 
hard to be parallelized when doing calculation numerically on parallel 
computers. For the Coulomb gauge method, it is open to question how much 
smearing the gauge fixing has done so the excited states contamination 
is not controlled. However, since the calculation after gauge fixing only 
needs the time direction links, its code is pretty easy to get parallelized. 
For example, when running on a 64-node QCDOC mother board the parallelized code 
takes about 4 minutes per configuration - about 40 times faster than the 
Bresenham code. We will see later that the finite size effects are much larger 
for the Coulomb than the Bresenham method. This will be discussed in the
next section.

Here I carried out a study of Coulomb gauge method on the
RBC-UKQCD 2+1 flavor domain wall fermion dynamical lattices 
with DBW2 and Iwasaki actions.
The work was done on $16^3\times 32$ as well as $24^3\times 64$
lattices. The simulation lattices and corresponding results are 
listed in the Table~\ref{tab:lattices}. I will compare most of 
them to the results from Bresenham method when corresponding data 
are available and see how consistent these two methods are. 
As the Coulomb gauge method costs much less computational power we 
can have a lot more statistics easily. In addition to this, we  
can further improve the statistics by choosing the "time" axis to 
lie along each of the four possible axes assuming the breaking 
of the $90^o$ rotation symmetry is negligible. We will confirm 
the symmetry breaking is small later in the results.

\begin{table}[ht]
    \begin{center}
    \begin{tabular}{|c|c|c|c|c|c|c|c|}
    \hline
    \hline
    $\beta$ & ($m_{ud},m_s$) &\# conf & $r_0/a$ (*)   & $a^{-1}$[Gev](*)& $r_0/a$(2) & $a^{-1}$[Gev](2)\\
    \hline
    \hline
    0.72    & (0.01,0.04) &  1000  & 3.962(30)& 1.564(12)& 3.947(38) & 1.558(15)  \\
    0.80    & (0.04,0.04) &  200   & 5.014(38)& 1.979(15)& 4.763(44) & 1.880(17) \\
    \hline
    \hline
    $\beta$/Vol & ($m_{ud},m_s$) &\# conf & $r_0/a$ (1)   & $a^{-1}$[Gev](1)& $r_0/a$(2) & $a^{-1}$[Gev](2)\\
    \hline
    \hline
           & (0.01,0.04)  & 600    & 3.997(22)  & 1.577(9)   & 4.011(21) & 1.583(8) \\
    2.13   & (0.02,0.04)  & 380    & 3.872(29)  & 1.528(11)   & 3.871(26) & 1.528(10)\\
    $16^3\times 32$ & (0.03,0.04)  & 300    & 3.815(30)  & 1.506(12)   & 3.825(28) & 1.510(11)\\
           & -$m_{res}$&-  & 4.113(31)  & 1.623(12)   & 4.127(44) & 1.629(17)\\
    \hline
           & (0.01,0.04) &  200    & 4.071(18)  & 1.607(7)    & 4.056(14) & 1.600(6)  \\
    2.13   & (0.02,0.04) &  200    & 3.946(12)  & 1.557(5)    & 3.947(11) & 1.558(5)  \\
    $24^3\times 64$ & (0.03,0.04) &  200    & 3.869(19)  & 1.527(7)    & 3.876(11) & 1.530(5)  \\
           & -$m_{res}$& - & 4.191(31)  & 1.654(12)    & 4.163(24) & 1.643(9)  \\
    \hline
    \end{tabular}
    \caption{Preliminary results for the DWF dynamical Lattices analyzed in this work. The first 2 ensembles have 
      DBW2 gauge action with volume $16^3\times 32\times 8$, and the $\beta=2.13$ ensembles have $Ls=16$ with 
      Iwasaki gauge action. (1) denotes the exponential fit results, (2) is for the constant fit results
      and (*) is from the Bresenham. The chiral limit results are extrapolated by taking $m_{ud}=-m_{res}
      =-0.00308$.}
    \end{center}
    \label{tab:lattices}
\end{table}

The potential can be determined by fitting the approximate
expression of the Wilson loop $\left\langle
W(\vec{r},t)\right\rangle=C(\vec{r})e^{-V(\vec{r})t}$. one can
either fit this using (1) ``exponential fit'': fit the Wilson loop
as an exponential function of $t$ to extract $V(\vec{r})$ or (2)
``constant fit'': fit the quantity $V(\vec{r},t)=\mbox{ln}[\left\langle
W(\vec{r},t)\right\rangle/\left\langle W(\vec{r},t+1)\right\rangle]$ to 
a constant in $t$. The double exponential fit is also an option if we 
take into account the lowest excited states contamination, but it is 
very unstable. All the analysis in this proceeding for the potential 
are either constant fit or exponential fit.

Having determined the potential, one can calculate the lattice scale by fitting the 
potential to the form: $V(\vec{r})= V_0 - \alpha/|\vec{r}|+\sigma|\vec{r}|$. 
Then the Sommer scale $r_0/a=\sqrt{(1.65-\alpha)/\sigma}$ is determined in 
lattice units. The lattice scale can be obtained by assuming the physical value 
of $r_0$ to be 0.5 fm.

\section{Finite size effect}
The Coulomb gauge method suffers from the enhanced finite size effect.
Quark states have images on the neighbor lattices due to the periodic 
boundary condition. As the Wilson loop calculation doesn't involve 
spacial links connecting the quark and anti-quark states, the images 
will have considerable contribution to the Wilson loops in the Coulomb method.
In contrast, the Bresenham method prevents the images in the wall from having 
much effect with explicit Wilson lines connecting specific pairs of Wilson
lines in the time-direction.

\section{Results}
\begin{figure}[ht]
    \hfill
    \begin{minipage}{0.45\textwidth}
        \hspace{-0.7cm}
         \vspace{-0.4cm}
        \epsfig{file=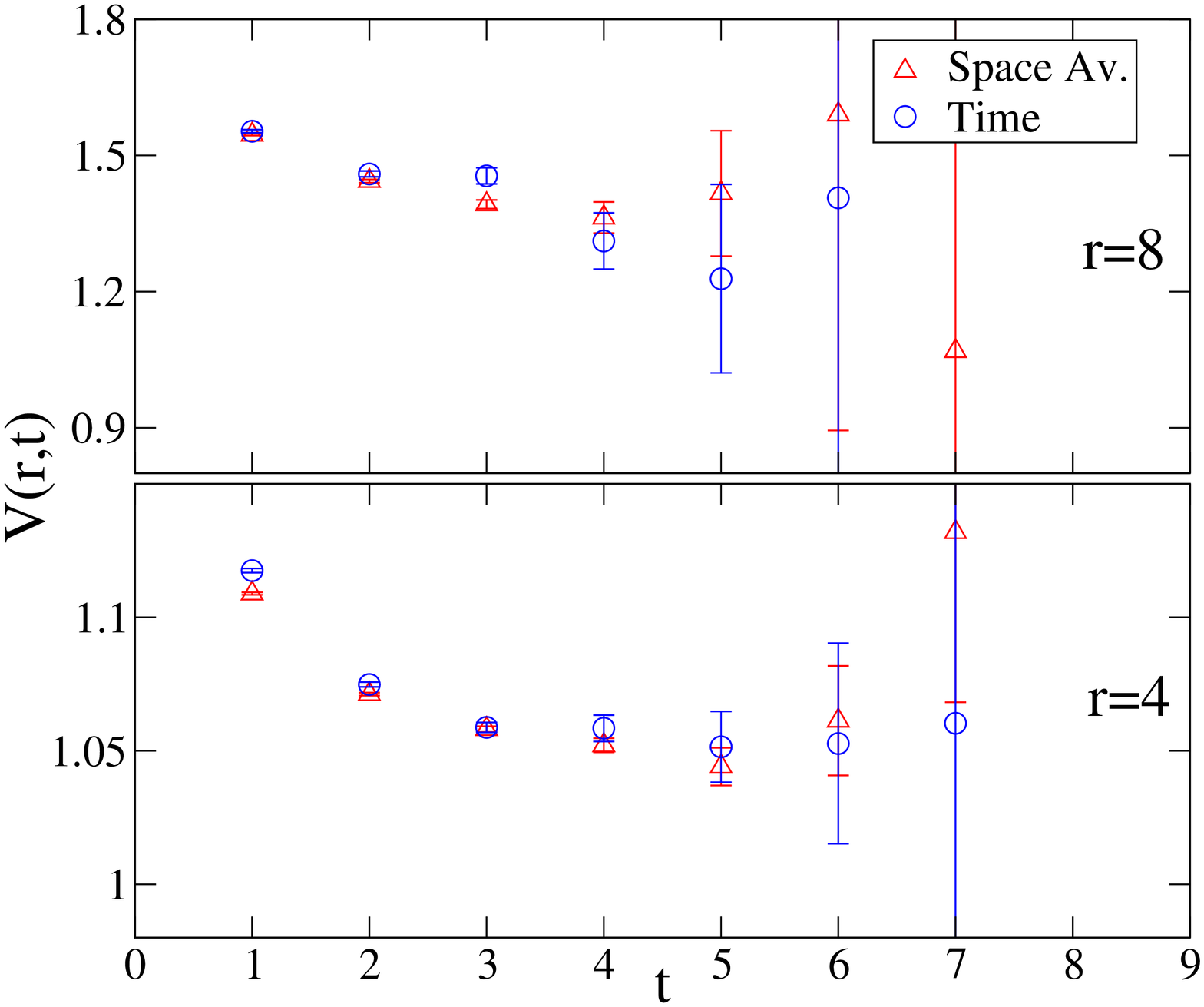,width=0.9\linewidth,angle=270} 
        \caption{The direction dependence of the potential for $\beta=2.13$, $16^3\times 32$: the spacial results averaged (triangle) and time results (circle).}
        \label{fig:DirDep}
    \end{minipage}
    \hfill
    \begin{minipage}{0.45\textwidth}
        \hspace{-0.7cm}
        \vspace{-0.3cm}
        \epsfig{file=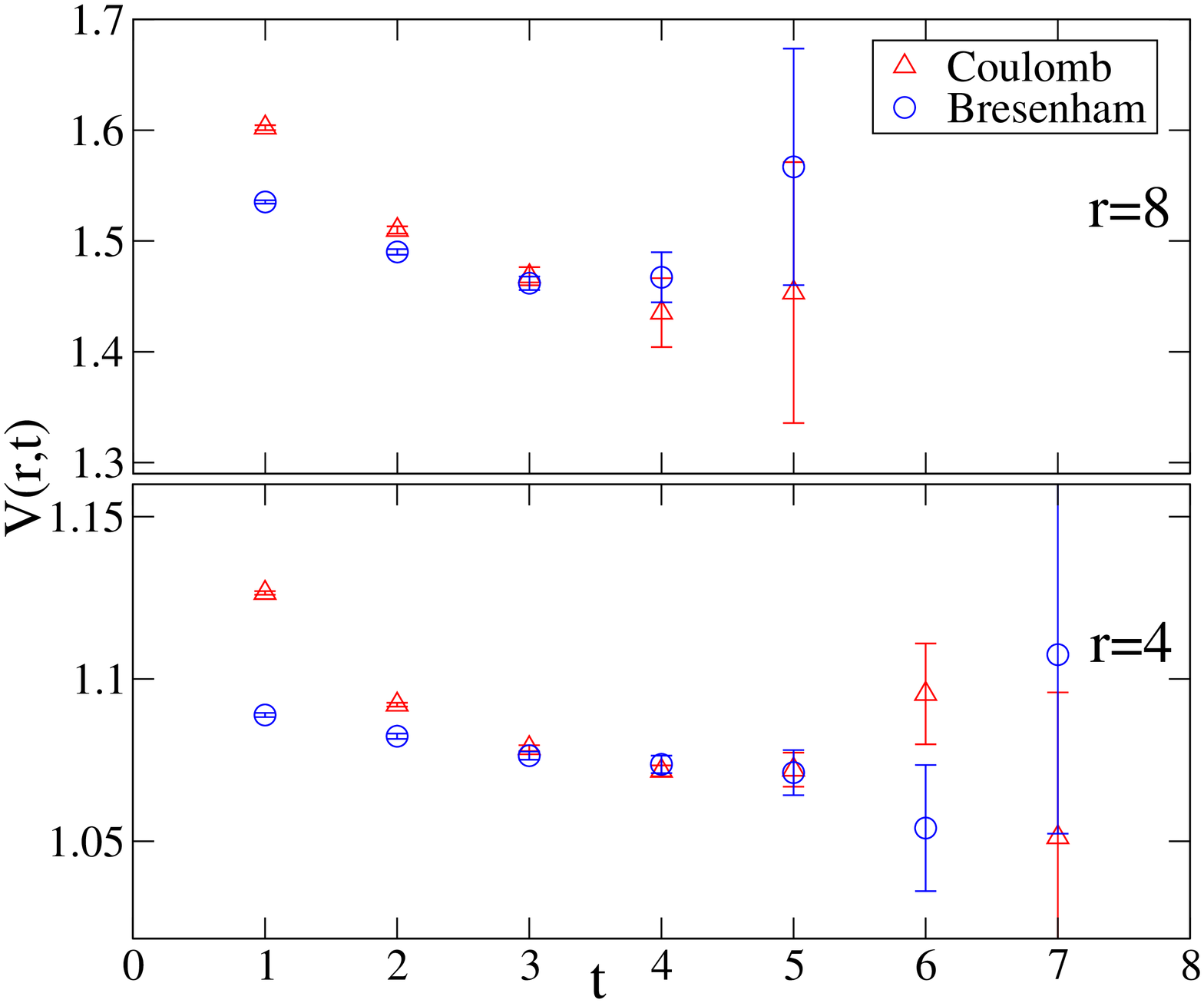,width=0.9\linewidth,angle=270}
        \caption{The time dependence of potential at fixed $r$'s for Coulomb (triangle) and Bresenham (circle)
          on $\beta=0.72$ ensemble.}
        \label{fig:0.72VT}
    \end{minipage}
    \hfill
\end{figure}

In this section, the Coulomb results on the potential and
$r_0$ will be compared to the Bresenham ones on all the 3 ensembles
- $\beta=0.72$, $\beta=0.80$ and $\beta=2.13$(Tab.~\ref{tab:lattices}).
Since we incorporated into the analysis our technique to increase
the statistics by 4 with "time" direction choices, we investigate 
how big is the rotational symmetry breaking. Fig.~\ref{fig:DirDep} 
is an example of the direction dependence of the potential on 
$\beta=2.13$, $16^3\times 32$, $m_{ud}=0.01$ lattices. The figure 
indicates the symmetry breaking is negligible. All the results 
for the Coulomb method are based on averaging potentials from 4 directions 
for each configuration. The Bresenham results are quoted from Ref.~\cite{koichi}
and are calculated with only one ``time'' axis. All errors are statistical only.

\subsection{$\beta=0.72$ ensemble}
As we use the constant fit, the potential obtained has dependence
on the time. And as the time increases, the excited states contamination
will decrease exponentially. Fig.~\ref{fig:0.72VT} shows the $t$
dependence of the potential at various $r$'s for the Bresenham and
Coulomb methods. The results presented here are based on 1000
configurations for the Coulomb method and 900 configurations for the 
Bresenham method. 
Both methods show good plateaus and the potential values are consistent
within the error bar when they reach the plateau.

\begin{figure}[ht]
  \vspace{-0.5cm}
    \hfill
    \begin{minipage}{0.45\textwidth}
        \hspace{-0.8cm}
        \vspace{-0.3cm}
        \epsfig{file=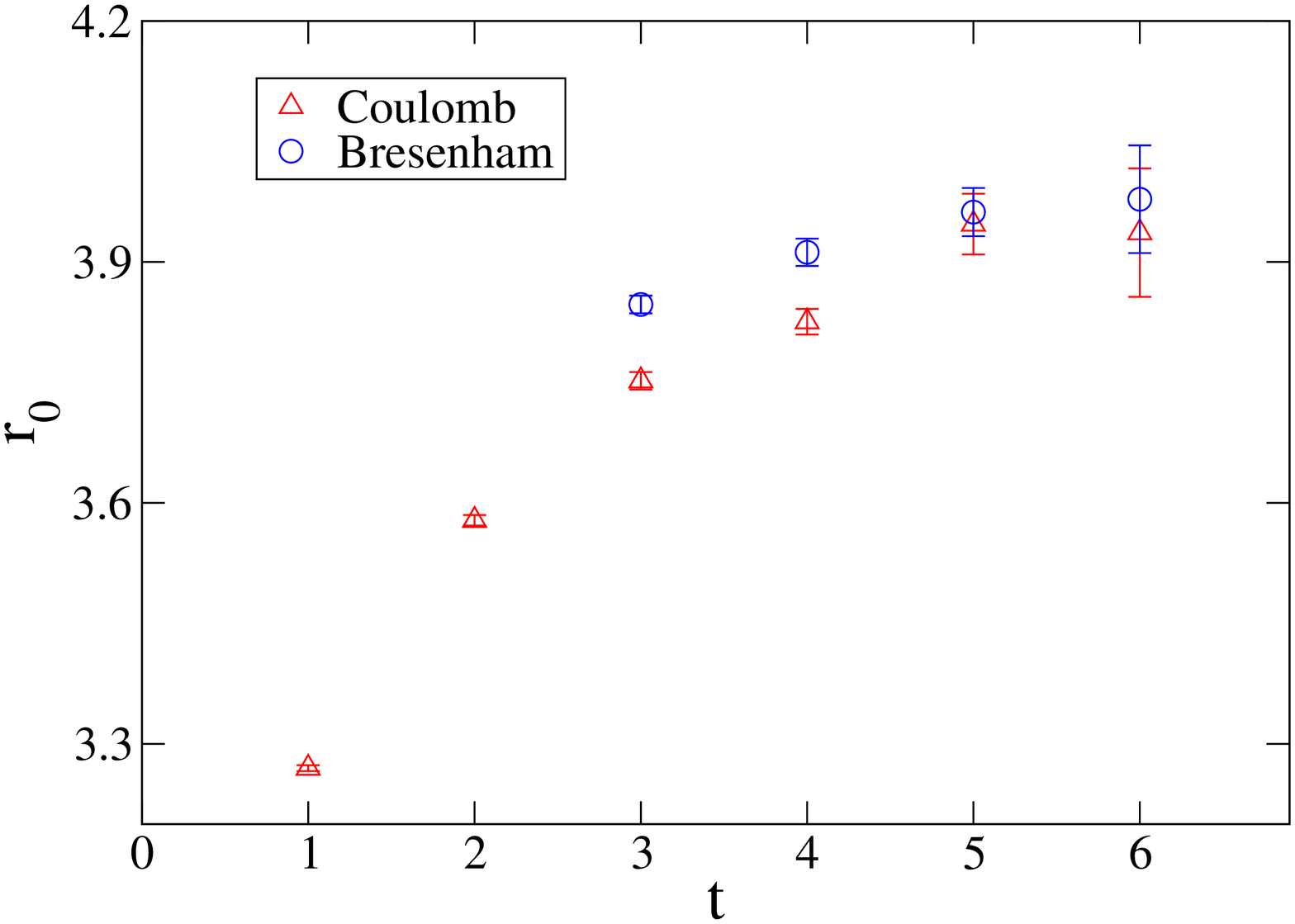,width=0.9\linewidth, angle=270} 
        \caption{The time dependence of $r_0$ for Coulomb (triangle) and 
	  Bresenham (circle) on $\beta=0.72$ ensemble.}
        \label{fig:r0-0.72}
    \end{minipage}
    \hfill
    \begin{minipage}{0.45\textwidth}
        \hspace{-0.7cm}
        \vspace{-0.3cm}
        \epsfig{file=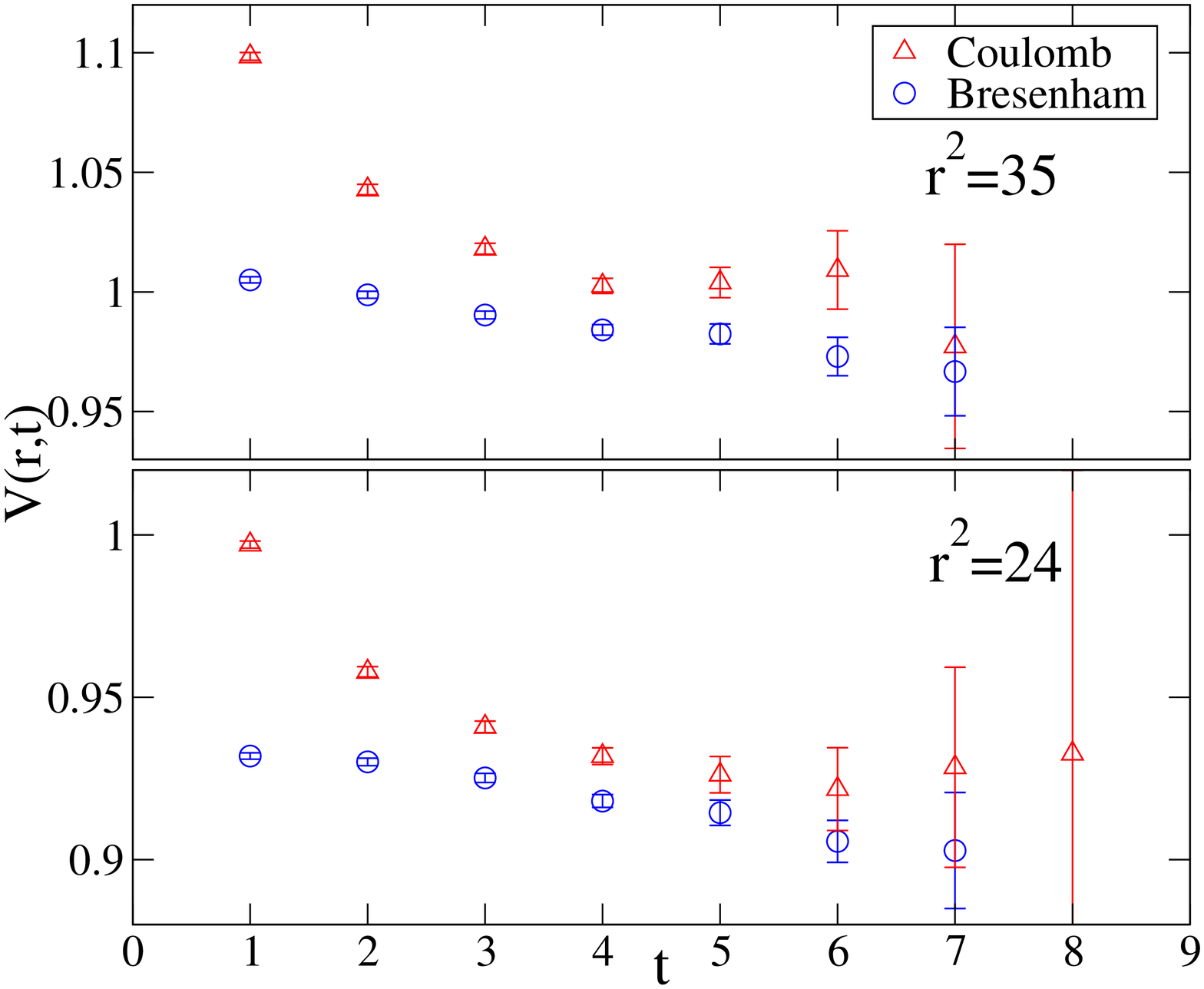,width=0.9\linewidth,angle=270}
        \caption{The time dependence of potential at some off-axis $r$'s for Coulomb (triangle) and Bresenham (circle) on $\beta=0.80$ ensemble.}
        \label{fig:0.80VT}
    \end{minipage}
    \hfill
\end{figure}

For $r_0$, we fit the potential as a function of $r$ at each
time slice. In order to get a sense of how different is Coulomb
from Bresenham, we choose the $r$ fitting range the same as in Ref.~\cite{koichi}: 
$\sqrt{3}$ to 8, the value quoted here is from $t=5$ only.  The comparison is given 
in Fig.~\ref{fig:r0-0.72}, which shows that (1) Both methods have
good plateaus indicating that the excited states contamination
is negligible; (2) $r_0$ values agree very well within error bars; 
(3) The errors are small(<1\%).

\subsection{$\beta=0.80$ ensemble}
\begin{figure}[ht]
  \vspace{-0.4cm}
    \hfill
    \begin{minipage}{0.45\textwidth}
        \hspace{-0.8cm}
        \vspace{-0.6cm}
        \epsfig{file=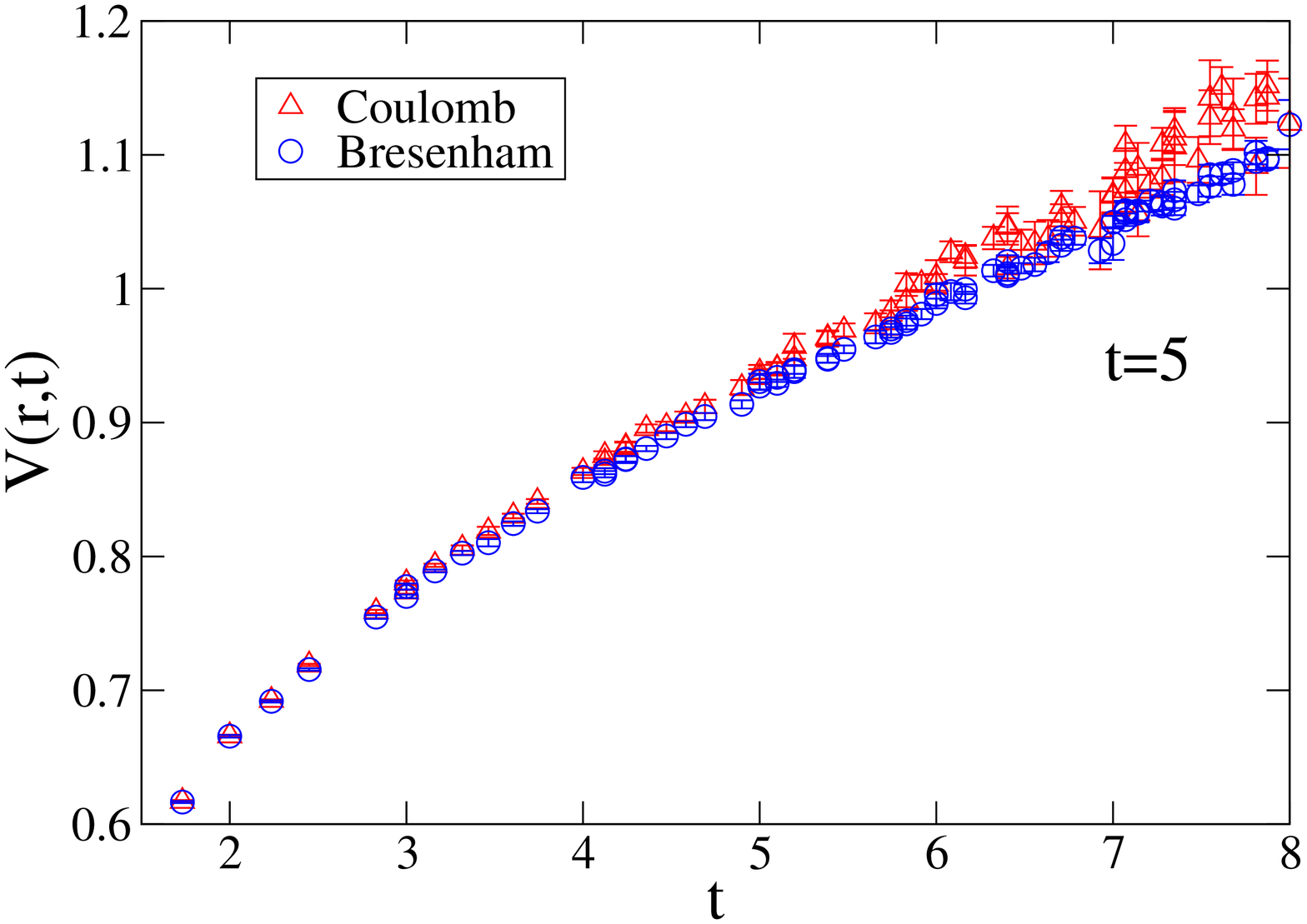,width=0.9\linewidth, angle=270} 
        \caption{The $r$ dependence of potential for Coulomb (triangle) and Bresenham (circle) on $\beta=0.80$ ensemble.}
        \label{fig:beta080PotT-5}
    \end{minipage}
    \hfill
    \begin{minipage}{0.45\textwidth}
        \hspace{-0.7cm}
        \vspace{-0.6cm}
        \epsfig{file=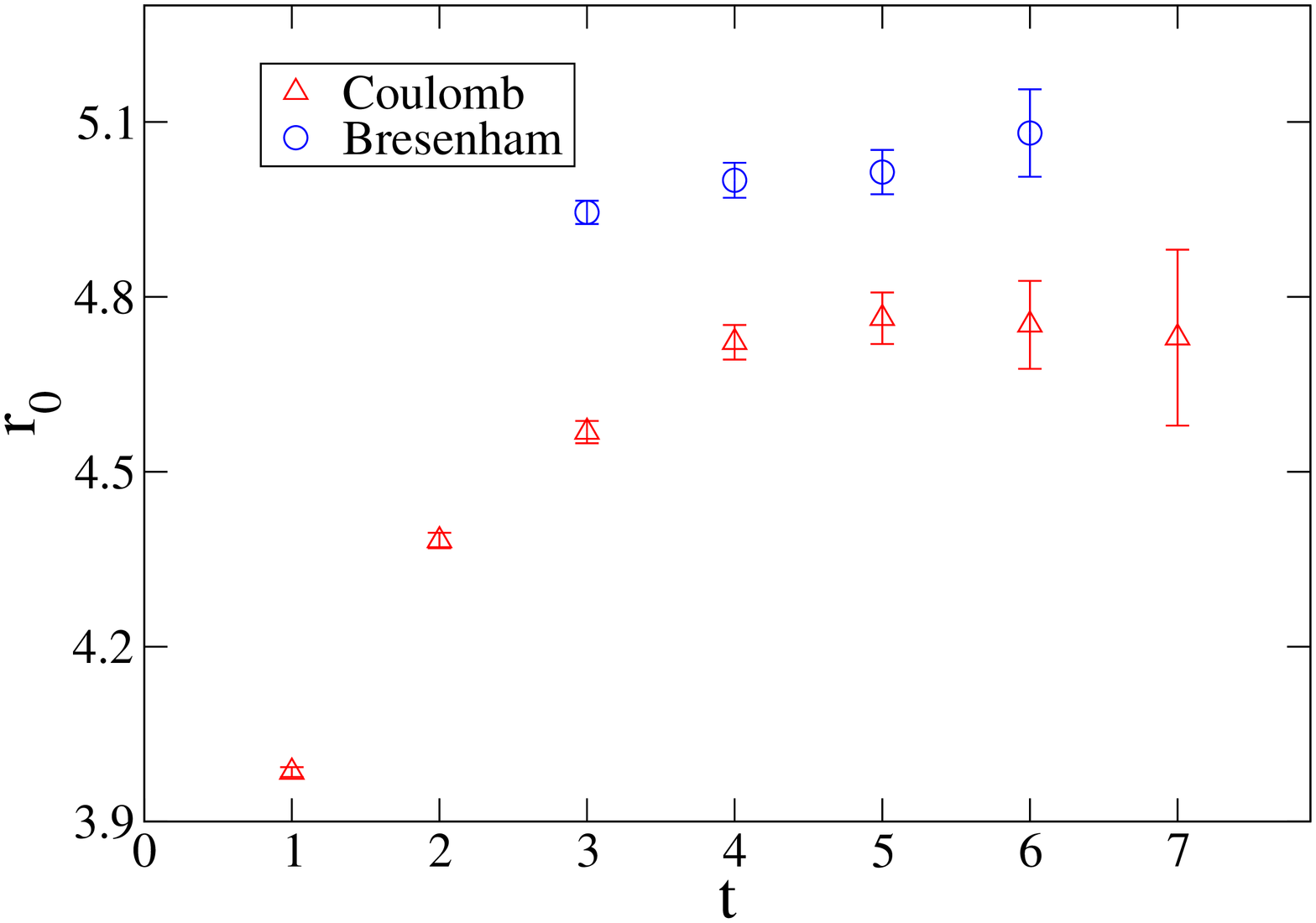,width=0.9\linewidth,angle=270}
        \caption{The time dependence of $r_0$ for Coulomb (triangle) and Bresenham (circle) on
        $\beta=0.80$ ensemble.}
        \label{fig:r0-080}
    \end{minipage}
    \hfill
\end{figure}

In this case, the finite size effect seems to have a large impact.
The results for both methods are based on the same 200 configurations.
To see a similar plot as in Fig.~\ref{fig:0.72VT} for $\beta=0.72$, we 
plot several off-axis $r$'s in Fig.~\ref{fig:0.80VT} as these manifest the potential
differences. The $r$ dependence of the potential is given in Fig.
~\ref{fig:beta080PotT-5} for $t=5$.

The fitted $r_0$ values at different times are plotted in
Fig.~\ref{fig:r0-080}. We get $r_0=4.763(44)$ for the Coulomb method and
$r_0=5.014(38)$ for the Bresenham method, both from $t=5$ only.  
There is a ~4-5\% discrepancy. By looking at the potential
[Fig.~\ref{fig:beta080PotT-5}], it seems consistent that the
finite size effect has much more impact on larger $r$.  And if we
shrink the fitting range to smaller $r$'s, one would expect the
potential for the two methods to become more and more consistent, thus 
the $r_0$ for the Coulomb method will also increase to match the
Bresenham. Although the $r_0$ and potential have large discrepancy
for this ensemble, it doesn't mean that the Coulomb method is
wrong and unreliable. To solve this problem, we can either use a
large volume lattice, or shrink the fitting range.

\subsection{$\beta=2.13$ ensemble}
\begin{figure}[ht]
    \hfill
    \begin{minipage}{0.45\textwidth}
        \hspace{-0.8cm}
        \vspace{-0.2cm}
        \epsfig{file=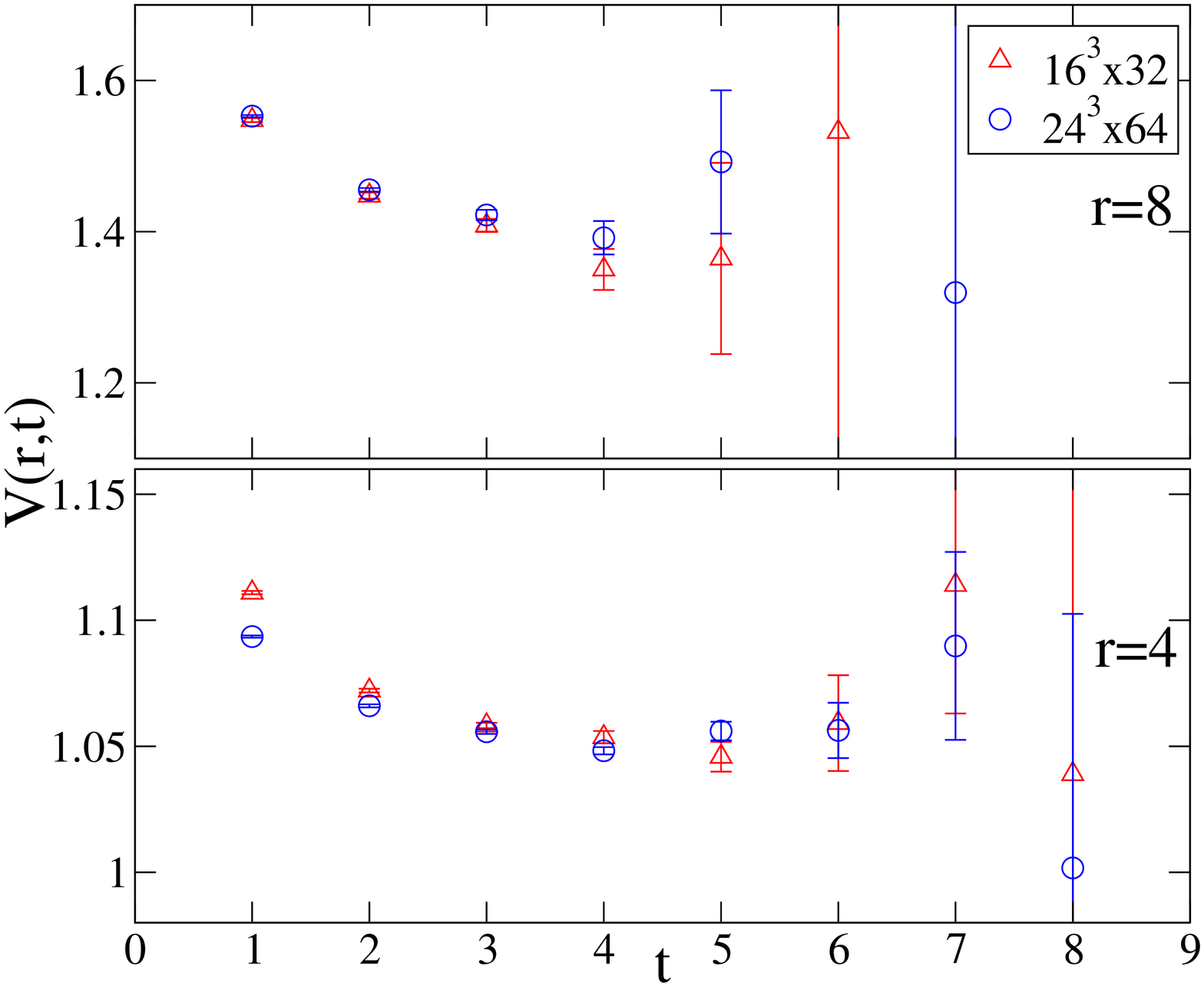,width=0.9\linewidth, angle=270} 
        \caption{The $r$ dependence of potential for Coulomb method on $\beta=2.13$, $m_{ud}=0.01$, $16^3\times 32$ and $24^3\times 64$ lattices.}
        \label{fig:VT-beta213}
    \end{minipage}
    \hfill
    \begin{minipage}{0.45\textwidth}
        \hspace{-0.7cm}
        \vspace{-0.2cm}
        \epsfig{file=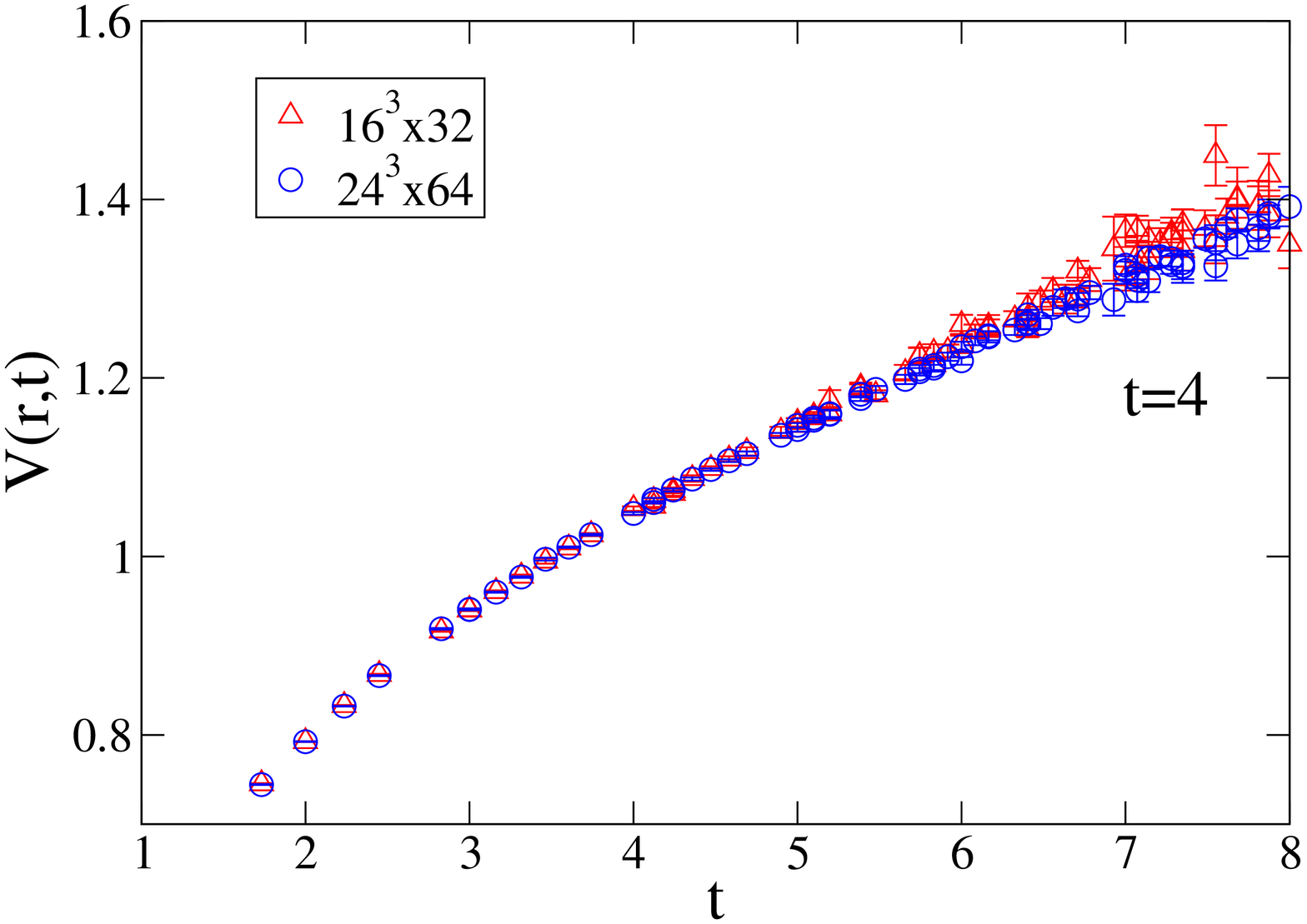,width=0.9\linewidth,angle=270}
        \caption{The potential comparison between $16^3\times 32$ and $24^3\times 64$ lattices
        for $\beta=2.13$ ensemble.}
        \label{fig:beta213PotT-5}
    \end{minipage}
    \hfill
\end{figure}

Here we have both $16^3\times32$ and $24^3\times64$ ensembles, and
for each ensemble 3 different sea quark masses are available so we
can do chiral extrapolation and set the lattice scale. The
comparison between $16^3\times 32$ and $24^3\times 64$ are plotted
for $m_{ud}=0.01$ case. From the potential plot
[Fig.~\ref{fig:VT-beta213},~\ref{fig:beta213PotT-5}] we can see
that the $24^3\times 64$ results are a bit lower at large r, which
is consistent with the proposed image problem. But since the
$Ls$ is different from $\beta=0.80$, there might be contamination
from other effects here. For the exponential fit, time range is from 4 to 6.
The constant fit results for $r_0$ are error weighted from t=4 to 6. 

\begin{figure}[ht]
  \vspace{-0.80cm}
    \hfill
    \begin{minipage}{0.30\textwidth}
        \hspace{-0.70cm}
        \epsfig{file=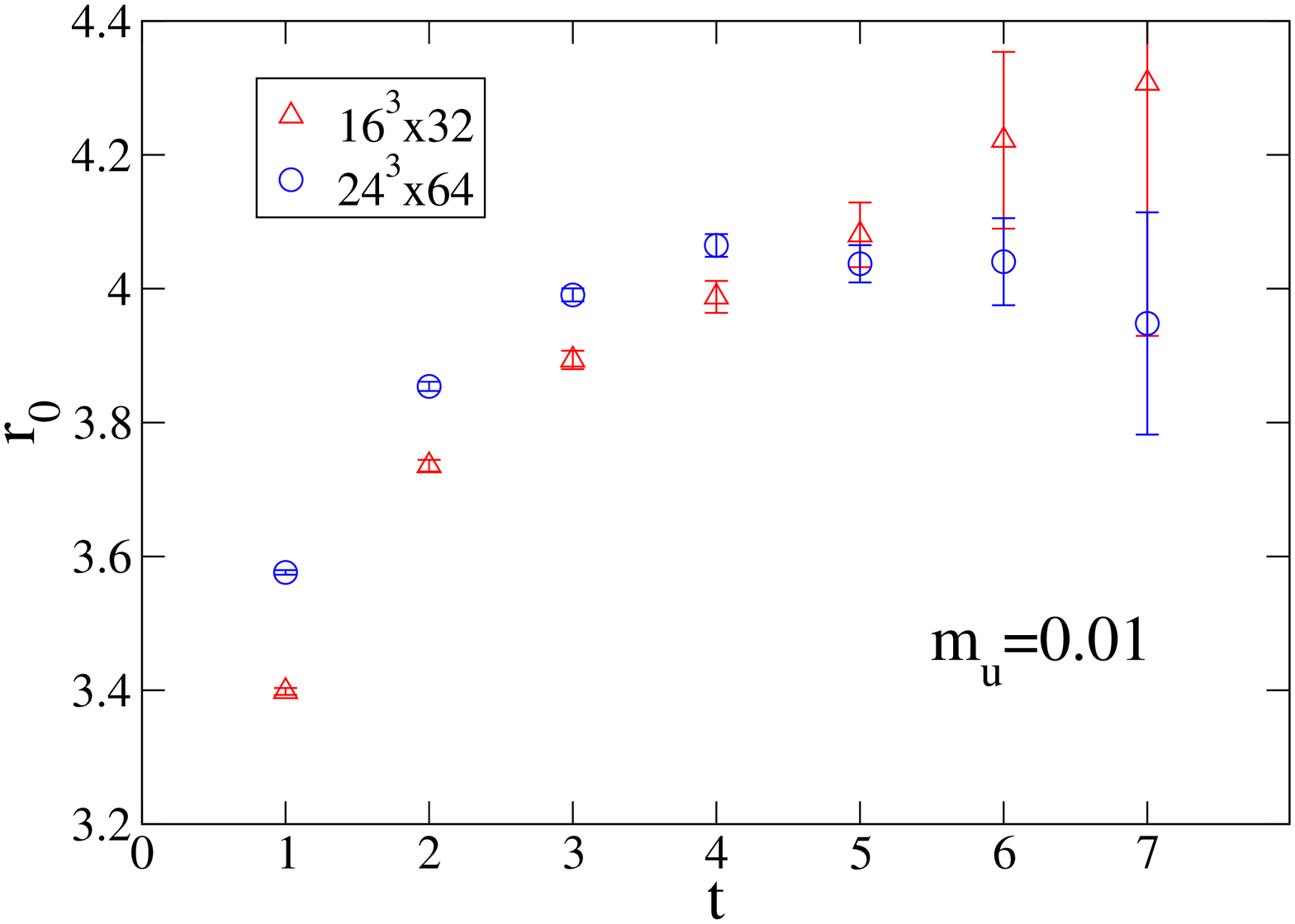,width=0.92\linewidth, angle=270} 
    \end{minipage}
    \hfill
    \begin{minipage}{0.30\textwidth}
        \hspace{-0.40cm}
        \epsfig{file=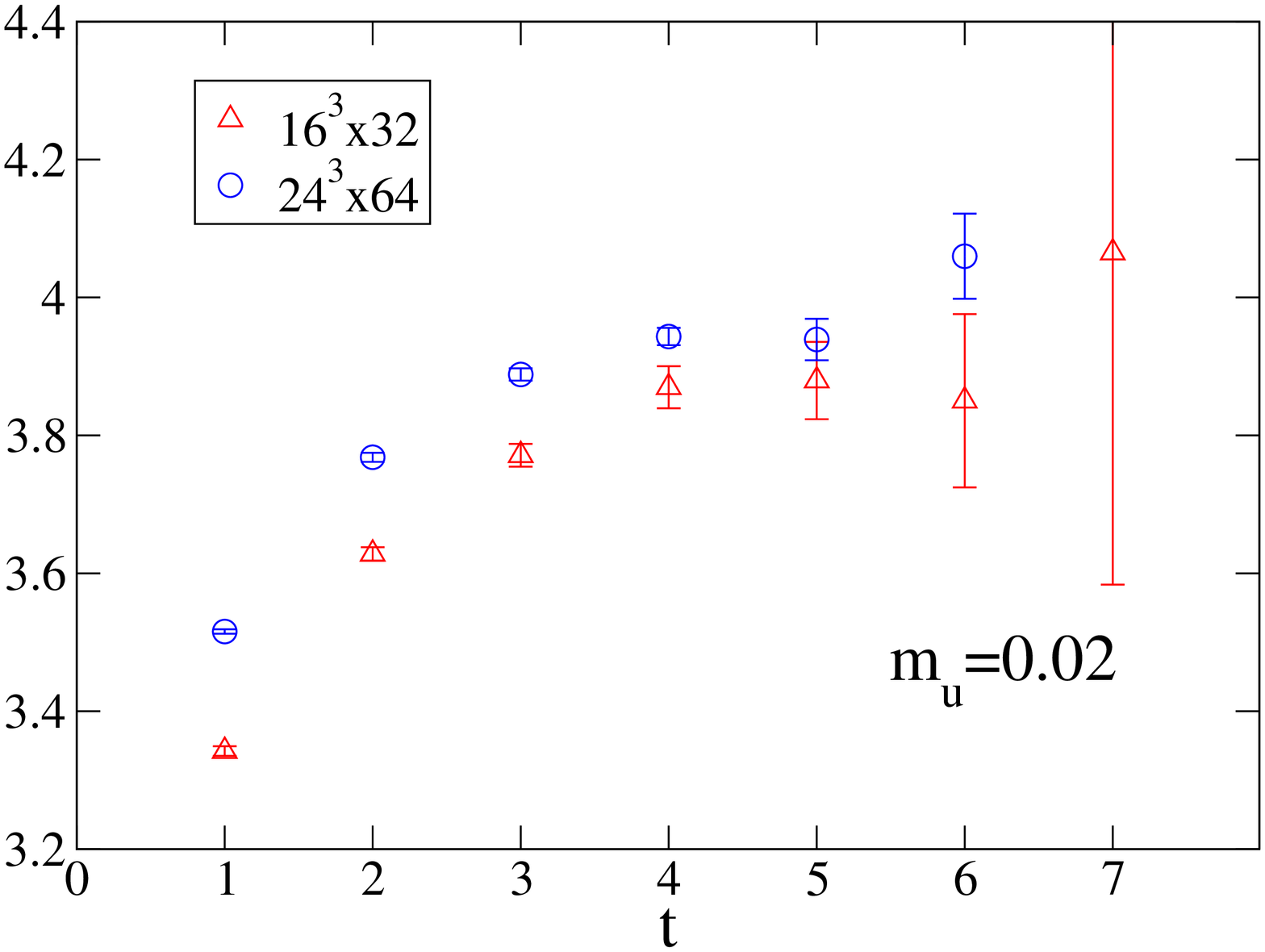,width=0.92\linewidth, angle=270} 
    \end{minipage}
    \hfill
    \begin{minipage}{0.30\textwidth}
        \hspace{-0.20cm}
        \epsfig{file=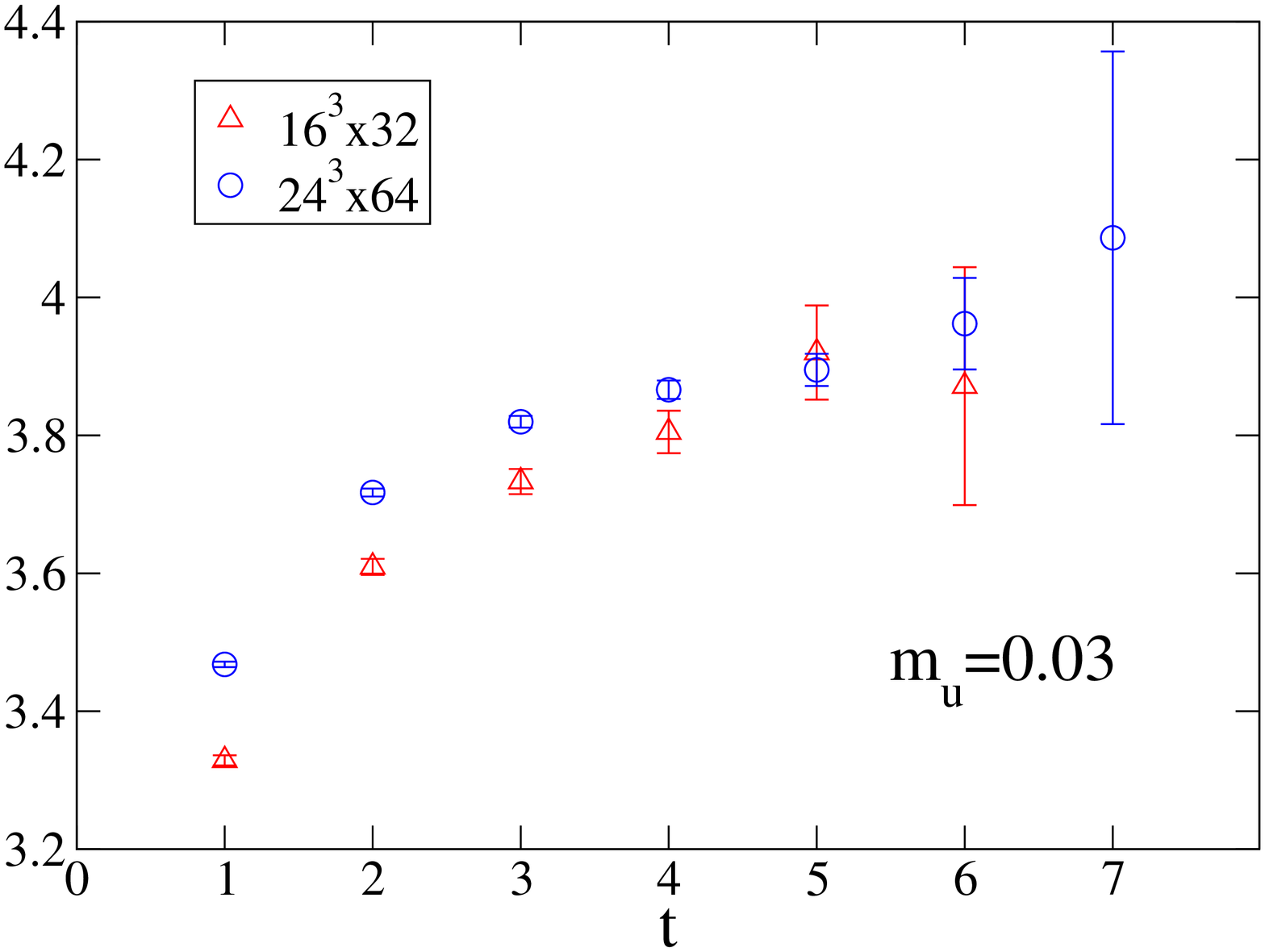,width=0.92\linewidth,angle=270}
    \end{minipage}
    \hfill
    \vspace{-0.5cm}
    \caption{The $r_0$ comparison between $16^3\times 32$ and $24^3\times 64$ lattices
        on $\beta=2.13$ ensemble.}
    \label{fig:r0-213}
\end{figure}

\begin{figure}[ht]
  \vspace{-0.6cm}
    \hfill
    \begin{minipage}[t]{0.45\textwidth}
        \hspace{-0.75cm}
        \epsfig{file=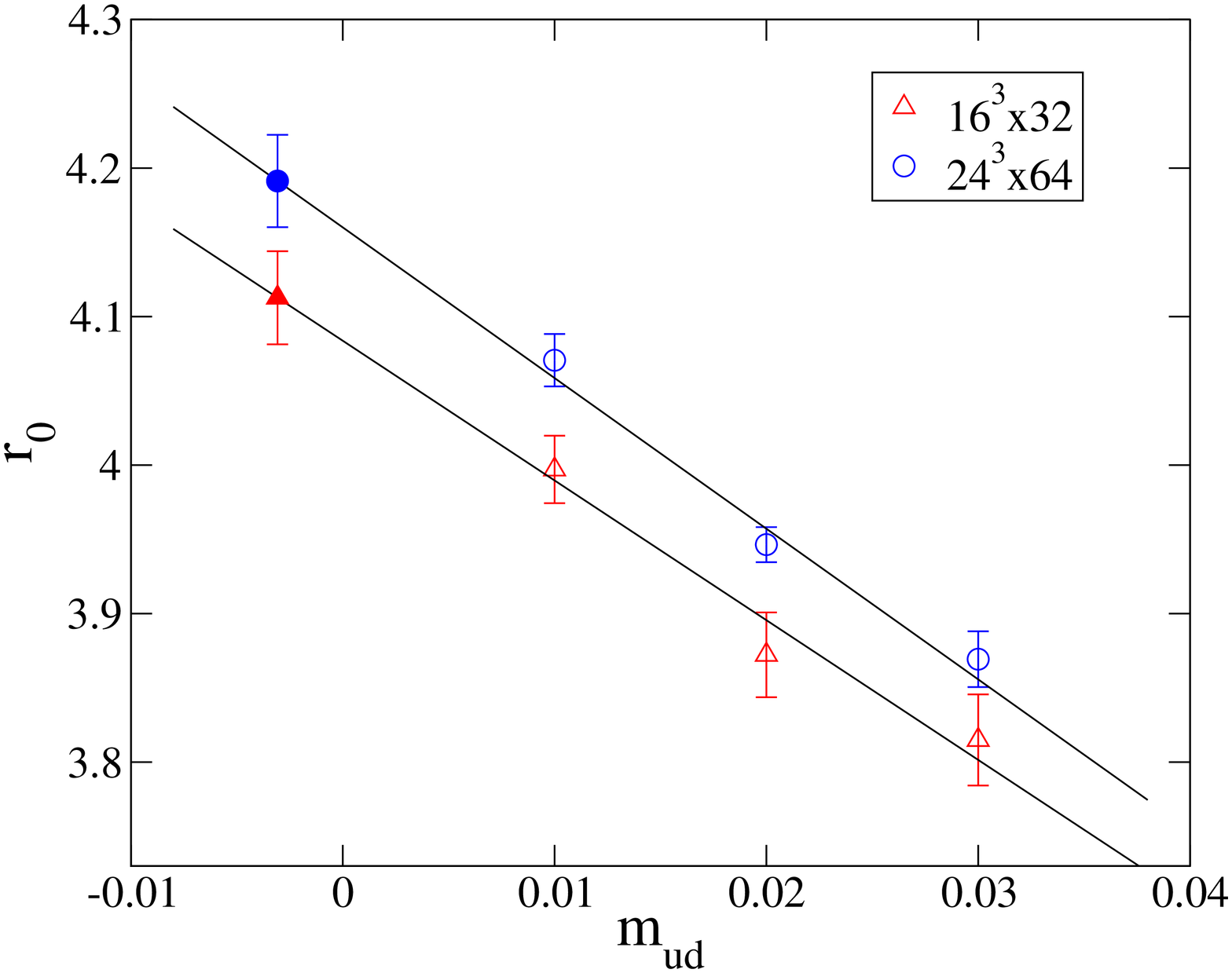,width=0.9\linewidth, angle=270} 
        \vspace{-0.75cm}
        \caption{Unitary chiral extrapolations for $r_0$ for $\beta=2.13$ ensemble on
        $16^3\times 32$ and $24^3\times 64$ lattices. Data are from exponential fit only.}
	\label{fig:extrap_16n24}
    \end{minipage}
    \hfill
    \begin{minipage}[t]{0.45\textwidth}
        \hspace{-0.75cm}
	\vspace{-0.5cm}
        \epsfig{file=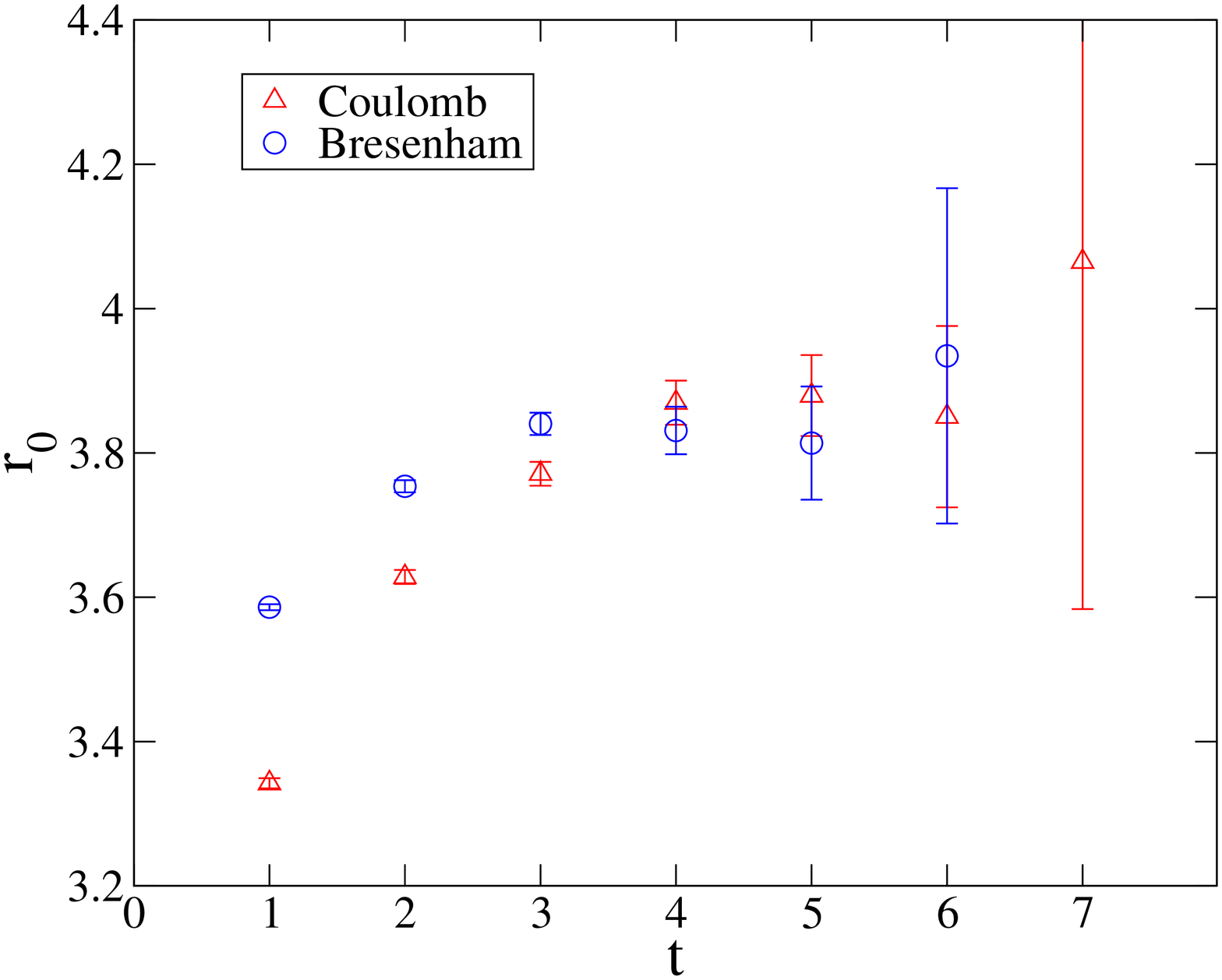,width=0.9\linewidth,angle=270}
        \caption{The Coulomb (triangle) and Bresenham (circle) results 
	  for $r_0$, on the $\beta=2.13$, $16^3\times 32$, $m_{ud}=0.02$ lattice.}
        \label{fig:cmp_b213_mu002}
    \end{minipage}
    \hfill
\end{figure}

The $r_0$ results from 3 different $m_{ud}$ are shown in Fig.~\ref{fig:r0-213}.
From the results of 2 volumes, unitary chiral extrapolation to the chiral limit give us 
the $r_0$ and the lattice scale $a^{-1}$, see Fig.~\ref{fig:extrap_16n24}. 
The $r_0$ values from $16^3\times 32$ are a bit lower. This is consistent with our observation on $\beta=0.80$.

Although the Bresenham results for the $\beta=2.13$ ensembles are not available in Ref~\cite{koichi}, 
I ran a parallelized QCDOC Bresenham code written by Takashi Umeda without tuning the
 APE smearing steps (fixed to 25) and produced some preliminary results for 
$\beta=2.13$, $16^3\times 32$, $m_{ud}=0.02$ case. As shown in Fig.~\ref{fig:cmp_b213_mu002}, 
the plateaus are good and the $r_0$ values are consistent. This indicates that the potential 
and lattice scale set by the Coulomb method is quite reliable.
 
\section{Summary and Conclusion}
We compare the Coulomb gauge method and the Bresenham method in different cases, 
and draw the conclusion that the Coulomb method is much easier to implement but
is more susceptible to finite size effects when the physical volume is small. We proposed 
the interaction with images as a possible explanation for the finite size effects. 
Although we haven't found a good way to eliminate the effects, a larger volume and a 
smaller fitting range for $r_0$ are candidates to make them negligible. The results 
on the $\beta=2.13$, $16^3\times 32$, $m_{ud}=0.02$ lattice for the Bresenham and 
Coulomb method are consistent. It indicates that the finite size effect doesn't contribute
much here so the chiral extrapolated $r_0$ and lattice scale set by this method should be 
reliable. And so does the $24^3\times 64$ case, in which the lattice has a much larger 
volume.

\section*{Acknowledgment}
We acknowledge helpful discussions with Norman Chirst, Taku Izubuchi and Robert Mawhinney, 
especially Taku Izubuchi for proposing the possible finite size effect problem. We thank Koichi Hashimoto
for his Bresenham data and Takashi Umeda for his parallelized QCDOC Bresenham code. In addition, 
we thank Peter Boyle, Dong Chen, Norman Christ, Mike Clark, Saul Cohen, Calin Cristian, Zhi-hua Dong,
Alan Gara, Andrew Jackson, Balint Joo, Chulwoo Jung, Richard Kenway, Changhoan Kim, Ludmila Levkova, 
Huey-Wen Lin, Xiaodong Liao, Guofeng Liu, Robert Mawhinney, Shigemi Ohta, Tilo Wettig and 
Azusa Yamaguchi for the development of QCDOC hardware and its software. The development and 
the resulting computer equipment were funded by the U.S. DOE grant DE-FG02-92ER40699, PPARC 
JIF grant PPA/J/S/1998/00756 and by RIKEN. This work was supported by U.S. DOE grant 
DE-FG02-92ER40699 and we thank RIKEN, BNL and the U.S. DOE for providing the facilities 
essential for this work.

\end{document}